\documentclass[fleqn,usenatbib]{mnras}

\usepackage[T1]{fontenc}

\DeclareRobustCommand{\VAN}[3]{#2}
\let\VANthebibliography\thebibliography
\def\thebibliography{\DeclareRobustCommand{\VAN}[3]{##3}\VANthebibliography}

\usepackage{graphicx}	
\usepackage{amsmath}	
\usepackage{amssymb}	

\usepackage{newtxtext,newtxmath}

\title[SN in TESS]{The Observable Supernova Rate in Galaxy-Galaxy Lensing Systems with the TESS Satellite}

\author[Holwerda et al.]{B. W. Holwerda$^{1}$, S. Knabel$^{1}$, R. C. Steele$^{1}$, L. Strolger$^{2}$, J. Kielkopf$^{1}$, A. Jacques$^{1,3}$, \and and W. Roemer$^{1}$  \\
$^{1}$ University of Louisville, Department of Physics and Astronomy, 102 Natural Science Building, 40292 KY Louisville, USA.\\
$^{2}$ Space Telescope Science Institute, 3700 San Martin Dr, Baltimore, MD 21218, USA \\ 
$^{3}$ NOIR Labs, 950 N. Cherry Ave., Tucson, AZ 85719, USA\\
}

\date{Accepted XXX. Received YYY; in original form ZZZ}

\pubyear{2020}

\begin{document}
\label{firstpage}
\pagerange{\pageref{firstpage}--\pageref{lastpage}}
\maketitle

\begin{abstract}
The Transiting Exoplanet Survey Satellite (TESS) is the latest observational effort to find exoplanets and map bright transient optical phenomena. Supernovae (SN) are particularly interesting as cosmological standard candles for cosmological distance measures. The limiting magnitude of TESS strongly constrains supernova detection to the very nearby Universe ($m \sim$ 19, $z<0.05$). We explore the possibility that more distant supernovae that are gravitationally lensed and magnified by a foreground galaxy can be detected by TESS, an opportunity to measure the time delay between light paths and constrain the Hubble constant independently. 
We estimate the rate of occurrence of such systems, assuming reasonable distributions of magnification, host dust attenuation and redshift. There are approximately 16 type Ia and 43 core-collapse SN (SNcc) expected to be observable with TESS each year, which translates to 18\% and 43\% chance of detection per year, respectively. Monitoring the largest collections of known strong galaxy-galaxy lenses from Petrillo et al., this translates into 0.6\% and 1.3\% chances of a SNIa and SNcc per year. The TESS all-sky detection rates are lower than those of the Zwicky Transient Facility (ZTF) and Vera Rubin Observatory (VRO). However, on the ecliptic poles, TESS performs almost as well as its all-sky search thanks to its continuous coverage: 2 and 4\% chance of an observed SN (Ia or cc) each year. These rates argue for timely processing of full-frame TESS imaging to facilitate follow-up and should motivate further searches for low-redshift lensing system. 
\end{abstract}




\section{\label{s:intro}Introduction}

The Transient Exoplanets Survey Satellite \citep[TESS,][]{Ricker15} is an outstanding tool for exploring transient phenomena, such as supernovae on cosmological distance scales, in its all-sky survey. However, the limiting depth of the 2-minute or 30-minute (integrated) cadence on each sector limits the volume of the Universe that can be probed using TESS. 

Fortunately, there are many galaxy-galaxy strong gravitational lenses that can magnify more distant supernovae. Traditionally, these have been found mostly in spectroscopic surveys such as the Sloan Digital Sky Survey \citep[SDSS][]{SDSS-DR1,SDSS-DR14} and the Galaxy and Mass Assembly (GAMA) survey \citep{Driver09,Liske15,Baldry18}. Because the signal from both the lens and more distant source galaxy are present in a single fiber spectrum, one can identify these as blended spectra as well as estimate the redshifts of both galaxies (lens and source). The clean selection through blended spectra has resulted in a very high confirmation rate for programs based on this technique: the SLACS, BELLS, SLACS4MASSES surveys \citep{slacs1,slacs2,slacs3,slacs4,slacs5,slacs6,slacs7,slacs8,slacs9,slacs10,slacs11,slacs12,slacs13} and ongoing searches using GAMA \citep[][]{Holwerda15,Knabel20}.

However, the selection function for these gravitational lenses is a convolution of the spectroscopic target selection and whether the fiber encloses the Einstein ring of the gravitational lens. Hence, while they select clean samples, the on-sky number of strong gravitational lenses is more difficult to estimate \citep[e.g.][]{Knabel20}.

Enter large-scale gravitational lens identification, either through citizen science such as the GalaxyZoo \citep{Lintott08} or machine learning techniques. The latter has been particularly successful using a training set generated using existing elliptical galaxies with added lensed arcs as the training set \citep{Petrillo17,Petrillo18,Li20c}. 
In their application using the (colour) images of the 1500 square degree Kilo-Degree Survey \citep[KiDS][]{de-Jong13,de-Jong15,de-Jong17}, \cite{Petrillo19} found 1300 such gravitational lenses and a similar number was discovered in the Dark Energy Survey \citep{Jacobs19,Huang20a,Huang20b}.

Interest in lensed supernovae and other bright transient phenomena is particularly high at the moment because the time differences between observed lensed supernova images (due to the different paths taken by light through the lensing galaxy) are sensitive to the expansion rate of the Universe. \cite{Refsdal64} proposed to measure the value of the Hubble
constant ($H_0$) from the time delays of multiply-imaged
SNe \cite[For excellent reviews, see][]{Treu16,Oguri19}. 

Since the discovery of a multiply-imaged supernova \cite[appropriately named Refsdal, see][]{Kelly14,Kelly15a,Kelly15b}, the feasibility of measuring the value of $H_0$ with this method became possible \citep{Vega-Ferrero18}. An independent $H_0$ measurement would be timely given the recent discrepancy of the $H_0$ value measured by the two dominant independent cosmological probes -- the cosmic microwave background
\citep{Planck-Collaboration18} and local distance ladders \citep{Beaton16, Riess16, Riess18, Riess19}.

The possibility of a strongly lensed supernova has now gone from a early possibility \citep{Wang00,Porciani00,Goobar02a,Goobar02b,Holz01,Kostrzewa-Rutkowska13} to a realistic prospect to be detected in statistical samples with modern survey cadence and sensitivity. 
Strongly lensed supernova in galaxy-galaxy lenses have been reported in iPTF, a precursor to the Zwicky Transient Factory \citep{Goobar17}, and in Pan-STARRS \citep{Quimby14}. 

Interest is now moving to how near-future observatories  \citep[e.g., \textit{JWST}][]{Petrushevska18}, VRO/LSST \citep{Liao18,Tu19} and other time-domain surveys \citep{Goldstein19}) can find a transient event in the source galaxy of a strong gravitational lens.
\cite{Shu18} explored the supernova rate (SNR) expected for the SLACS and related programs (BELLS and SLACS4MASSES), given their spectroscopically selected lenses and their source redshift distribution. They show that facilities such as the Dark Energy Survey (DES) telescope and Vera Rubin Observatory (VRO) can very likely identify lensed supernovae candidates in these surveys. 

Our interest is to explore how well TESS would be able to detect such transients, either by monitoring known lenses or as a blind survey. Despite the fact that TESS limiting depths are much shallower than those considered in \cite{Shu18}, the on-sky density of lensing galaxies from Petrillo's result indicates that there could be a useful number of supernova detections via lensed sources observed in the current and future TESS mission. The high cadence and 27 day long light curves of TESS open the possibility of observing the SNIa light curve as it occurs in multiple images of the lensed source. The continuous viewing zone for TESS at both Ecliptic poles offers even better time coverage with 357 day coverage. Our aim in this paper is to explore how much of a scientific return can be expected by building an immediate search for supernova in the TESS data for rapid follow-up.

\begin{figure}
  \begin{center}
	\includegraphics[width=0.5\textwidth]{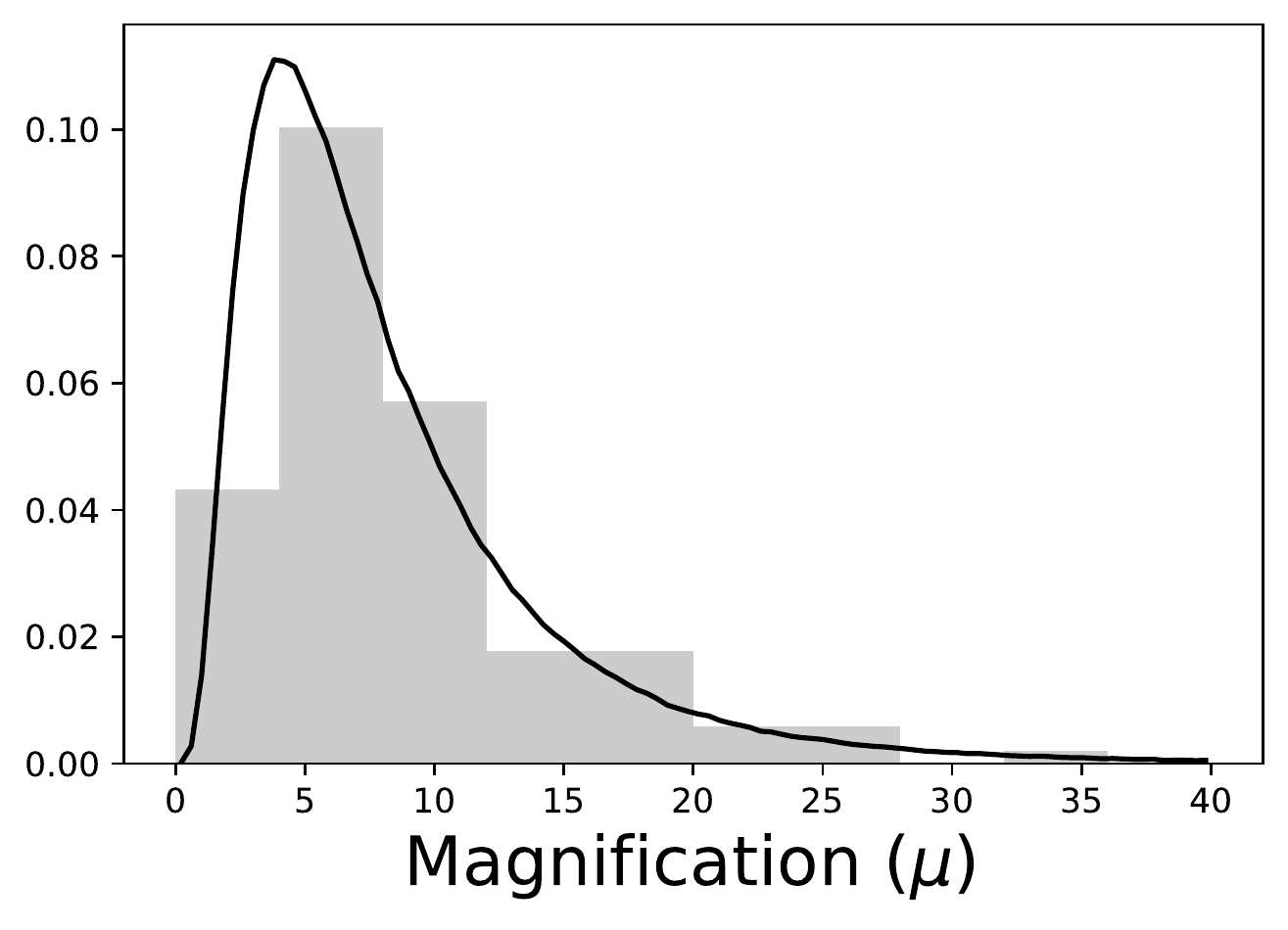}
	\caption{\label{f:magnification} The distribution of magnification by the foreground lensing galaxy of the background source we assume for our estimates. The histogram is the distribution of values in SLACS and the line best description lognormal distribution we use for the estimates. }
  \end{center} 
\end{figure}

\begin{figure}
  \begin{center}
	\includegraphics[width=0.5\textwidth]{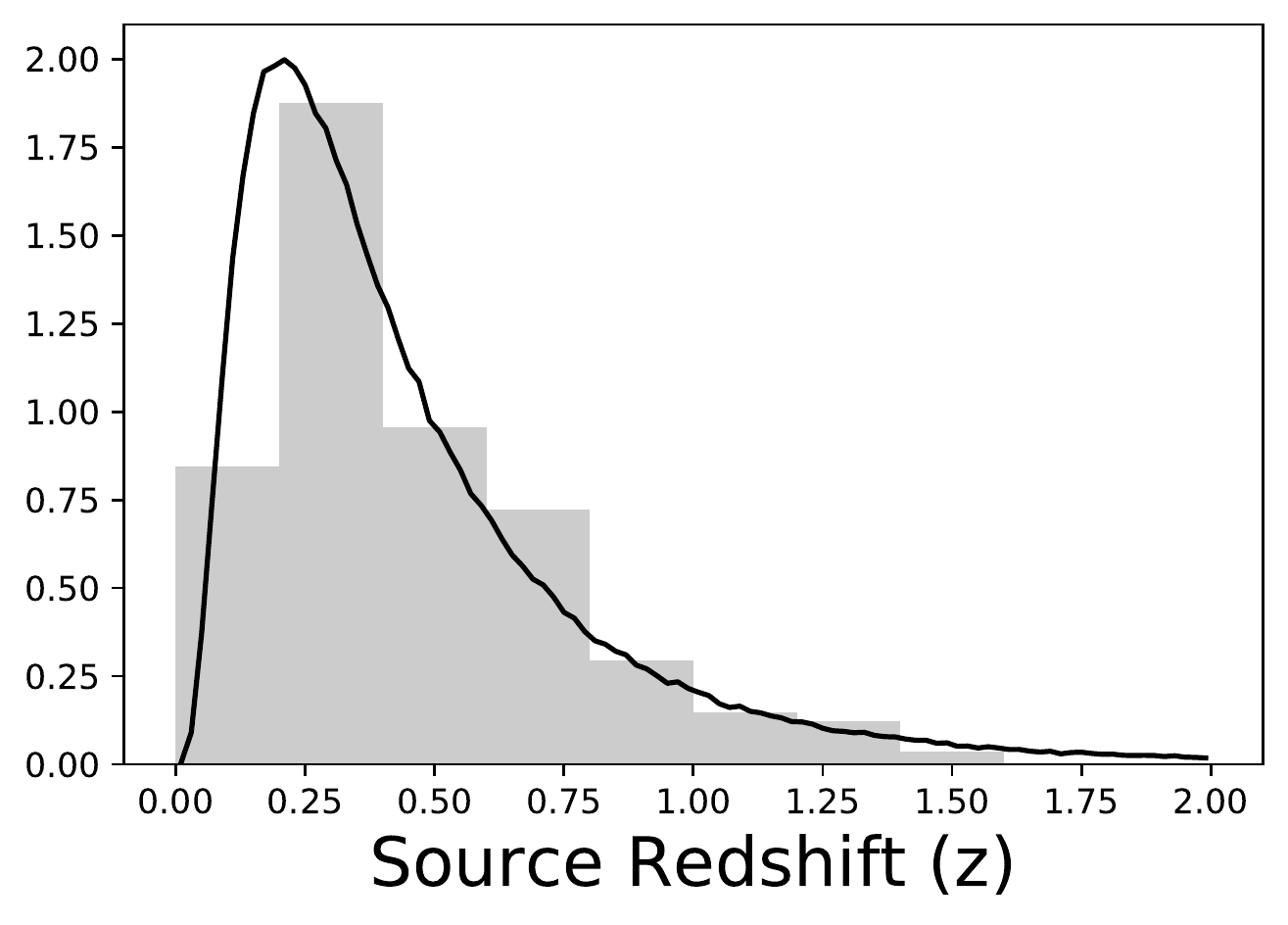}
	\caption{\label{f:z}The distribution of source galaxy redshifts assumed for our estimates. This is the combined redshift numbers from SLACS and GAMA (gray histogram) and the best lognormal distribution describing it (solid line).}
  \end{center} 
\end{figure}

\begin{figure}
  \begin{center}
	\includegraphics[width=0.5\textwidth]{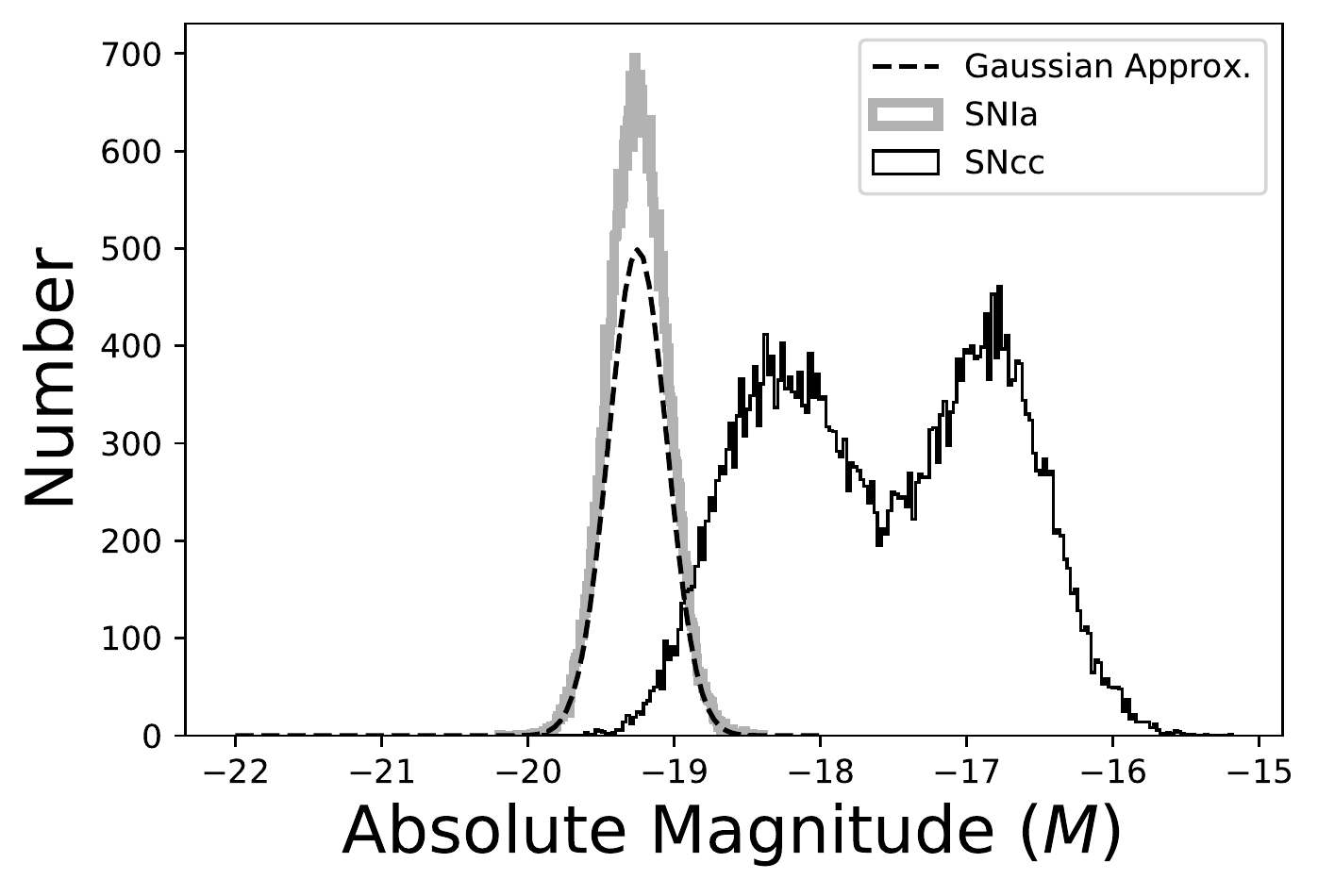}
	\caption{\label{f:absolute-magnitude}The absolute magnitude distributions of SNIa and core-collapse (Type IIb,IIL,IIP, and IIn combined) based on the values reported in \protect\cite{Richardson14}. }
	\end{center} 
\end{figure}

\begin{figure}
  \begin{center}
	\includegraphics[width=0.5\textwidth]{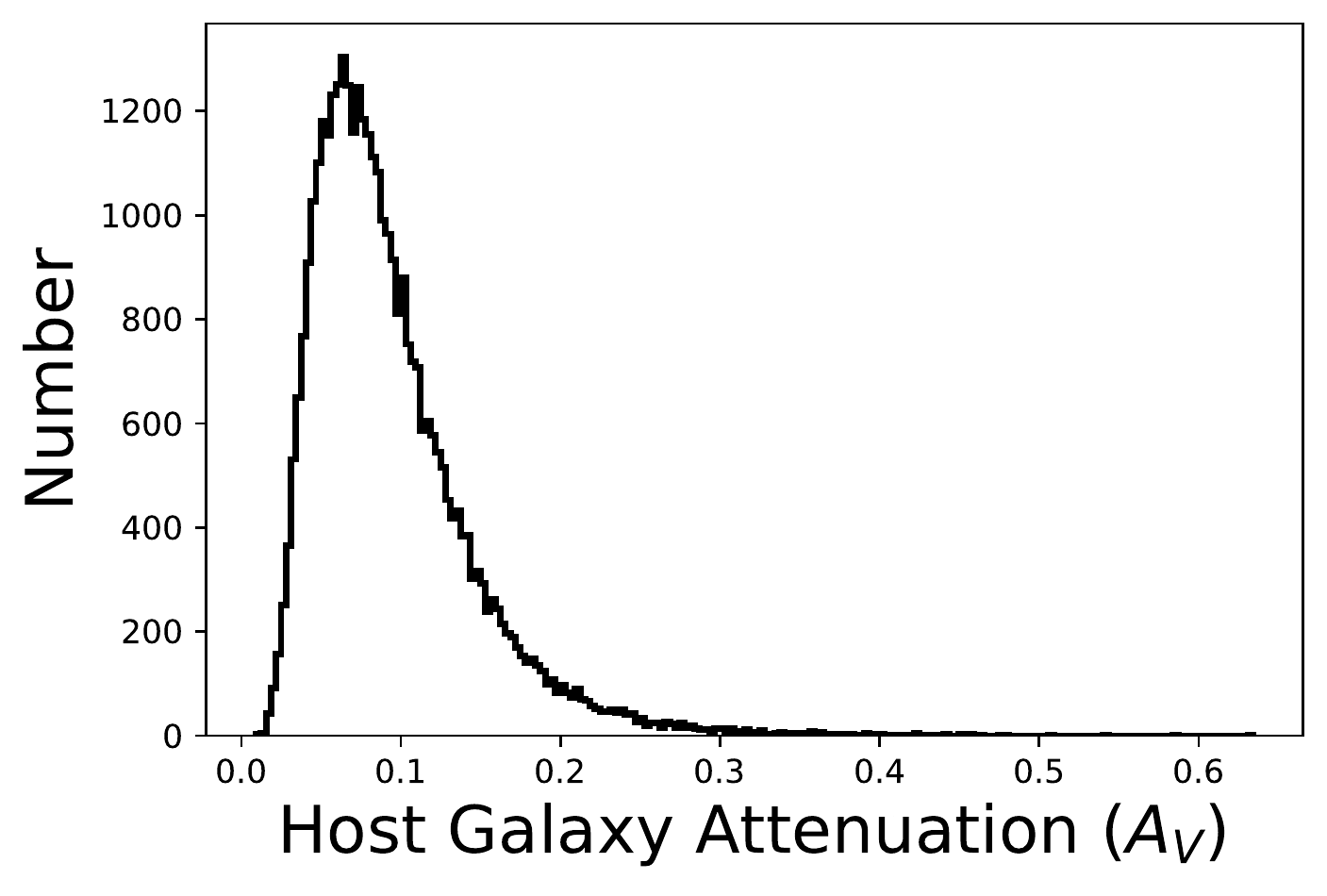}
	\caption{\label{f:host-attenuation}The 
	host galaxy attenuation distribution for a $10^9 M_\odot$ stellar mass disk galaxy from \protect\cite{Holwerda15c} based on the HST imaging presented in \protect\cite{Holwerda09}. Given the selection criteria for lens selection (blue arcs, implying smaller, star-forming galaxies), we adopt this distribution to randomly draw host attenuation ($A_V$) values from for each supernova.} 
	\end{center} 
\end{figure}

\begin{figure}
  \begin{center}
	\includegraphics[width=0.5\textwidth]{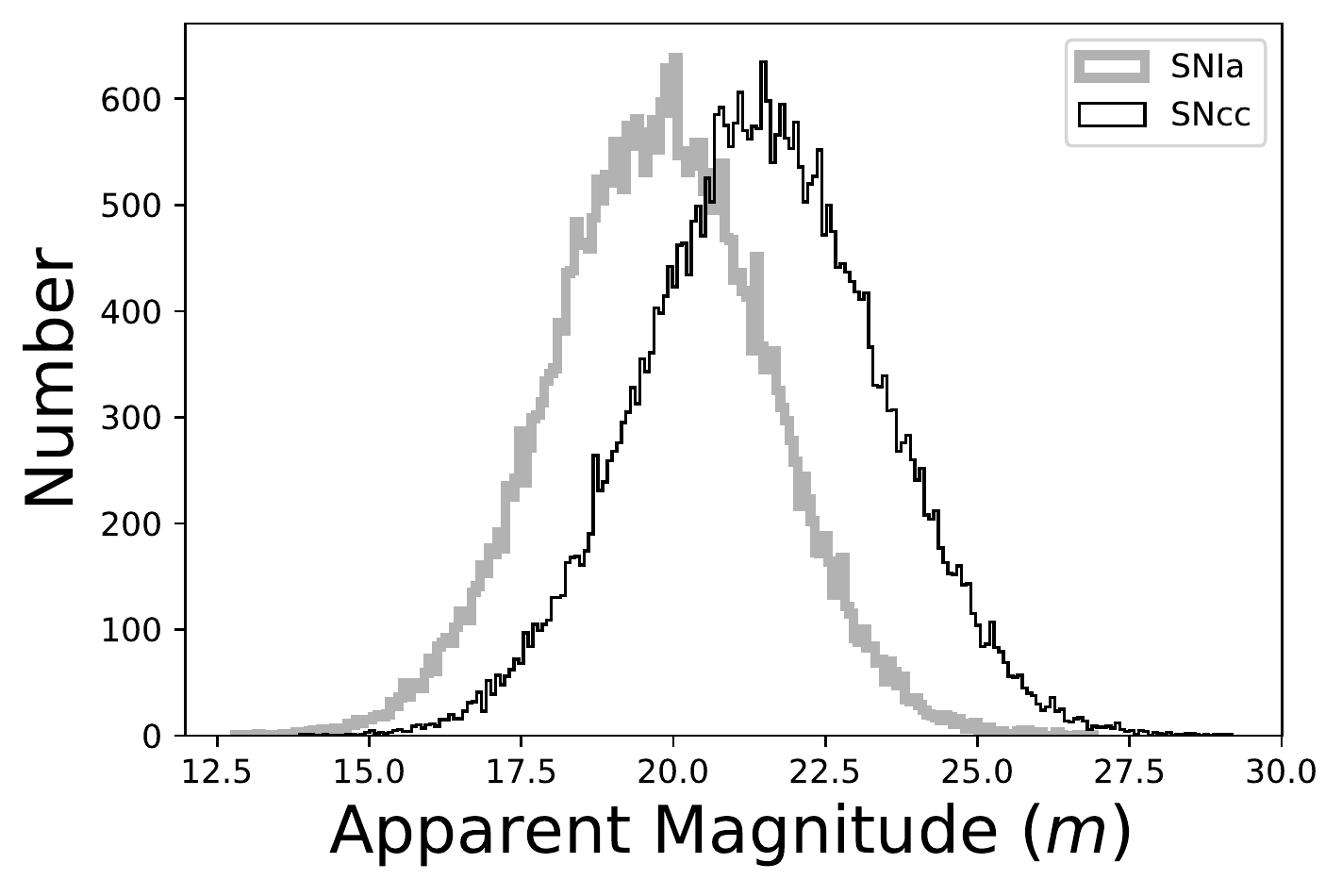}
	\caption{\label{f:apparent-magnitude}The apparent magnitude distributions of SNIa and core-collapse (Type IIb,IIL,IIP, and IIn combined) after randomly drawing from their absolute magnitude distribution and the redshift and magnification distributions in Figure \ref{f:magnification} and \ref{f:z} and adding host attenuation from Figure \ref{f:host-attenuation}. }
	\end{center} 
\end{figure}

\section{Estimating Lensed SN Rates in TESS}

We follow the same method demonstrated in \cite{Shu18}, populating the sky with lenses like those found in the Petrillo and other samples, and estimating the supernova rate, the apparent magnitude based on the \cite{Shu18} magnification distribution (Figure \ref{f:magnification}), the absolute magnitude distribution from \cite[][]{Richardson14} shown in Figure \ref{f:absolute-magnitude} and an assumed redshift distribution (Figure \ref{f:z}) based on the source distribution in \cite{Holwerda15} and \cite{Shu18} combined. 
We follow the approach in \cite{Shu18} for the TESS estimates by populating $N$ lenses using either supernova type Ia absolute magnitudes and the four core-collapse supernovae types listed in \cite{Richardson14}\footnote{\cite{Li11b} has the more complete fractions of observed types (their Figure 11). For SNe Ia, we focus on the most numerous normal Type Ia, and disregarding rarer and less luminous 91bg and 02cx types.}. 
Drawing from the normal distribution defined by the mean and standard deviation of absolute magnitudes in Table 1 of \cite[][]{Richardson14} and equally randomly drawing from the four distributions for core-collapse supernovae, we obtain the distributions in Figure \ref{f:absolute-magnitude}.

The magnification distribution in Figure \ref{f:magnification} is from \cite{Shu18} alone, as this is the most uniform sample available. We fit a lognormal distribution ($\mu=1.9$, $\sigma = 0.7$) to this distribution, which is a reasonable description (K-S=0.1, $p=0.05$) using the Kolmogorov-Smirnov test; it deviates a maximum of 10\% from the lognormal description.

The redshift distribution in Figure \ref{f:z} is drawn from the combination of source redshifts from \cite{Shu18}, \cite{Holwerda15} and \cite{Knabel20}. The source redshift distribution from \cite{Holwerda15} follows a lognormal distribution (K-S=0.1, p=0.003), as does the one from \cite{Shu18} but with much lower significance (K-S=0.04, p=0.98). The combined sample is reasonably described with a lognormal distribution as well (K-S = 0.06, $p = 0.07$). We adopt the lognormal approximation for the source redshifts ($\mu =-1.04$, $\sigma = 0.75$) to describe the redshift distribution, as it deviates from the lognormal distribution by less than 10\%. 

As volume increases at higher redshifts, the chances of alignment and strong lensing increase. However, we assume here that the redshift distribution is appropriate for TESS detected lensed supernovae as (a) we assume Petrillo-like lenses in the blind survey and (b) the Petrillo ML algorithm was trained on SLACS-like artificial lenses. This assumption does ignore higher redshift ($z>0.5$) lenses, lensing high-redshift sources ($z\sim1$) with extreme magnifications. As a result, our estimates for the TESS detected supernovae rates are likely slightly underestimated. \cite{Knabel20} estimated the on-sky density of lenses such as those in SLACS but the combined selection effects will miss strongly lensed events and the estimate of all-sky lensed events presented later are under-estimates. 

To estimate the host galaxy attenuation, we adopt the distribution of $A_V$ values from \cite{Holwerda15c}, shown in Figure \ref{f:host-attenuation}, which is based on the overlapping galaxy pair originally described in \cite{Holwerda09,Holwerda13b}. From overlapping pairs of galaxies \citep{kw00a,kw00b,kw01a,kw01b,Holwerda07c,Holwerda13a,Holwerda16}, we know there is a wide variety of attenuation distributions depending on stellar mass and positions in the disk of a galaxy. This particular template is for a $10^9 M_\odot$ stellar mass disk galaxy, which is a reasonably choice for the star-forming galaxies preferred for the source galaxies (lenses are identified in colour images from blue arcs). In effect, host galaxy is both inclination dependent \citep{Holwerda15a} and strongly depends on the relative distribution of dust and SN progenitors \citep[][Holwerda+ {\em in prep}]{Holwerda08a}.

We now combine the absolute magnitude distribution with the luminosity distances for each redshift, the magnifications and the host attenuation distribution to obtain apparent magnitude distributions for both Type Ia supernovae and core-collapse supernovae. This is effectively convolving the distributions in Figure \ref{f:absolute-magnitude} with the distributions in Figures \ref{f:magnification} and \ref{f:z} to arrive at the apparent magnitude distribution in Figure \ref{f:apparent-magnitude}. These distributions are close to Gaussian with $m = 19.69 \pm 1.78$ and $m=21.38 \pm 1.94$ for SNIa and SNcc respectively; these mean values fall below a reasonable TESS limit but have a wide enough spread to potentially be observable. 

The on-sky density of strong lenses needs to be estimated as well and this number is not yet well settled. From the initial pass by \cite{Petrillo19}, there is about 0.8 lens per square degree. However, \cite{Knabel20} estimate a higher on-sky density based on a mix of identification techniques, closer to 1.27 lens per square degree by combining all three identification methods. Even so, this number is likely to be an under-estimate.  Using the latter as our on-sky density with 70\% of the area accessible by TESS (due to the Zone of Avoidance and Zodaical light), we estimate the approximate number of strong lenses in the TESS survey to be $\sim37$ thousand lenses. Starting with an absolute magnitude (Figure \ref{f:absolute-magnitude}), and randomly drawing from the redshift and magnification distributions (Figures \ref{f:magnification} and \ref{f:z}), we arrive at the distribution of apparent magnitudes in Figure \ref{f:apparent-magnitude} in the all-sky lens survey.
The redshifts assigned to each random draw can be translated into a SNR for each type using the relation found by \cite{Shu18}. 

\begin{figure*}
  \begin{center}
  	\includegraphics[width=0.49\textwidth]{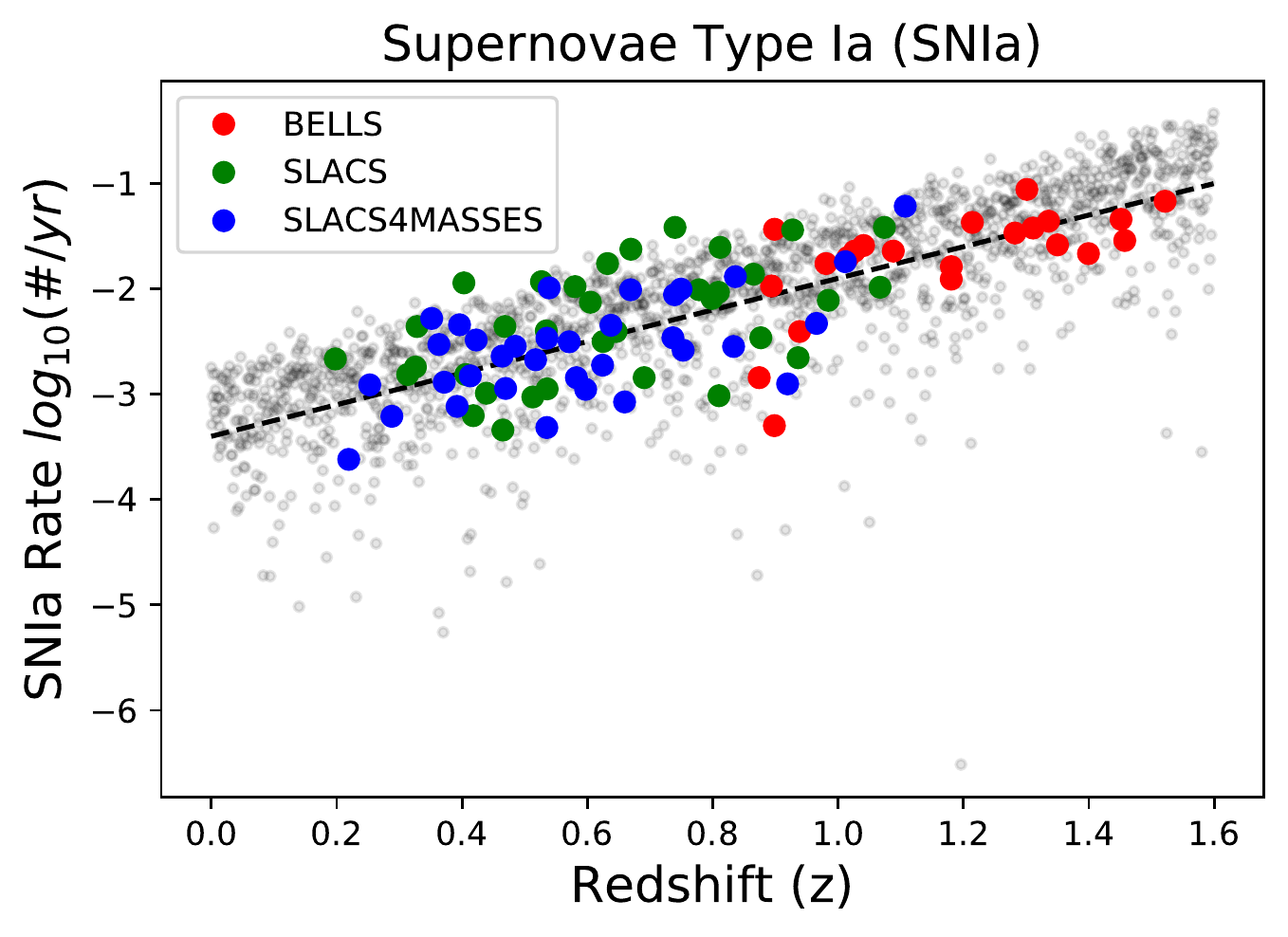}
  	\includegraphics[width=0.49\textwidth]{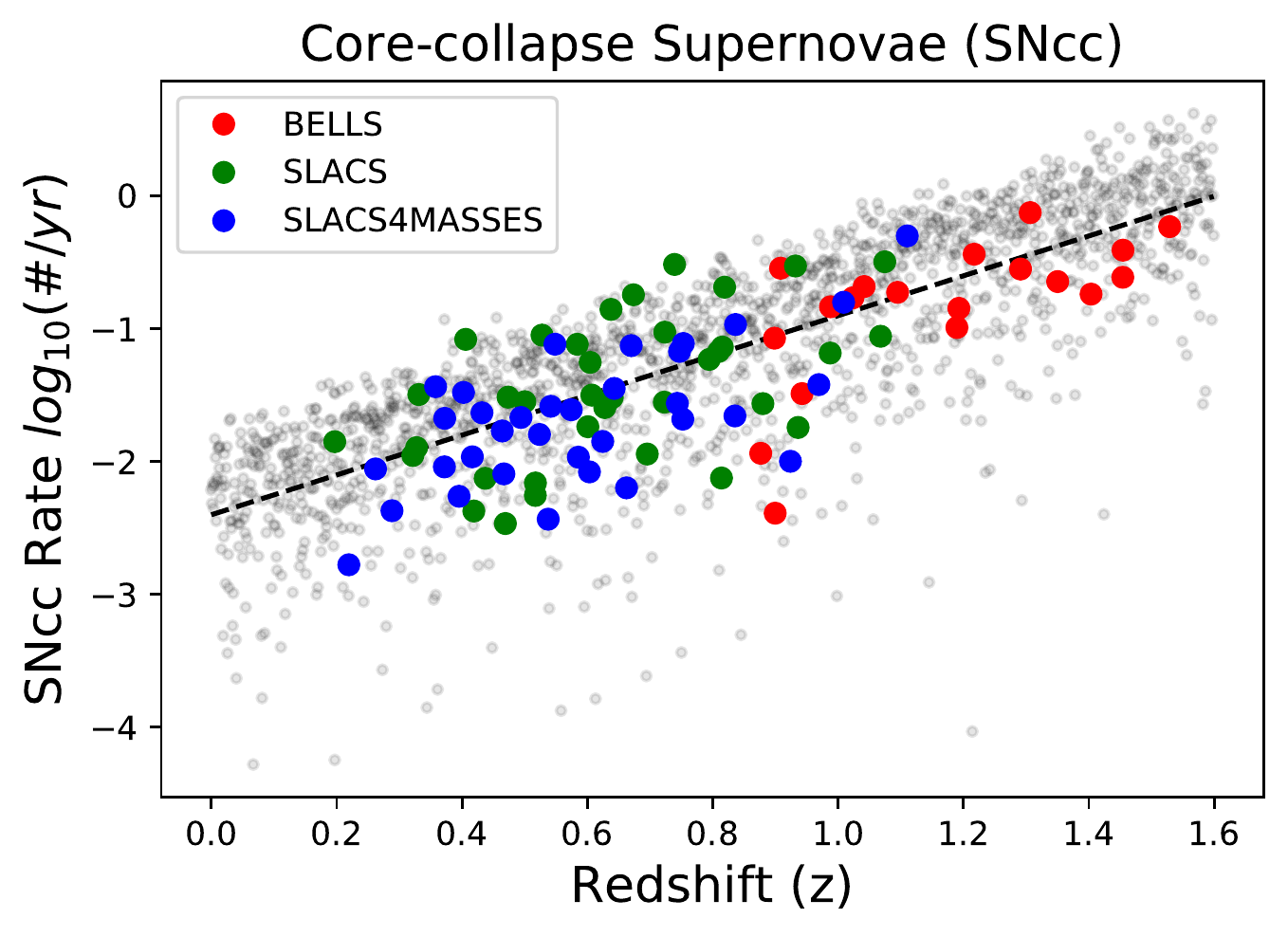}
	\caption{\label{f:snr:z}The relation between supernova rates and redshift for SNIa (left) and core-collapse SN (right). The estimated SNIa and SNcc rates from \protect\cite{Shu18} for the three strong lensing surveys are shown. The dashed line is the best linear fit through these values. We use this fit and the inferred scatter around it to populate the lenses drawn for the TESS survey.}
  \end{center} 
\end{figure*}

\cite{Shu18} estimated the star-formation rates in each lensed galaxy in SLACS using the [OII] emission line. The core-collapse supernova rate is  directly proportional to the recent SFR \citep{Dahlen99,Oguri10} with a factor for galaxy mass \citep[see][]{Strolger15}. A similar reasoning can be followed for the SNIa rate \citep{Dahlen99}. We fit a linear relation to the values in the \cite{Shu18} lenses (colour points in Figure \ref{f:snr:z}) and populate the all-sky survey based on the drawn redshift and the variance of SNR at that redshift (gray points in Figure \ref{f:snr:z}). Please note that this implicitly uses the redshift distribution shown in Figure \ref{f:z} for the rates, and that likely higher rates at greater redshift are not fully included in the all-sky estimate.

The supernova rates for both SNIa and SNcc are inferred from the spectroscopic star-formation rate by \cite{Shu18} for the source galaxies as a function of redshift (Figure \ref{f:snr:z}).

\begin{figure*}
  \begin{center}
  	\includegraphics[width=0.49\textwidth]{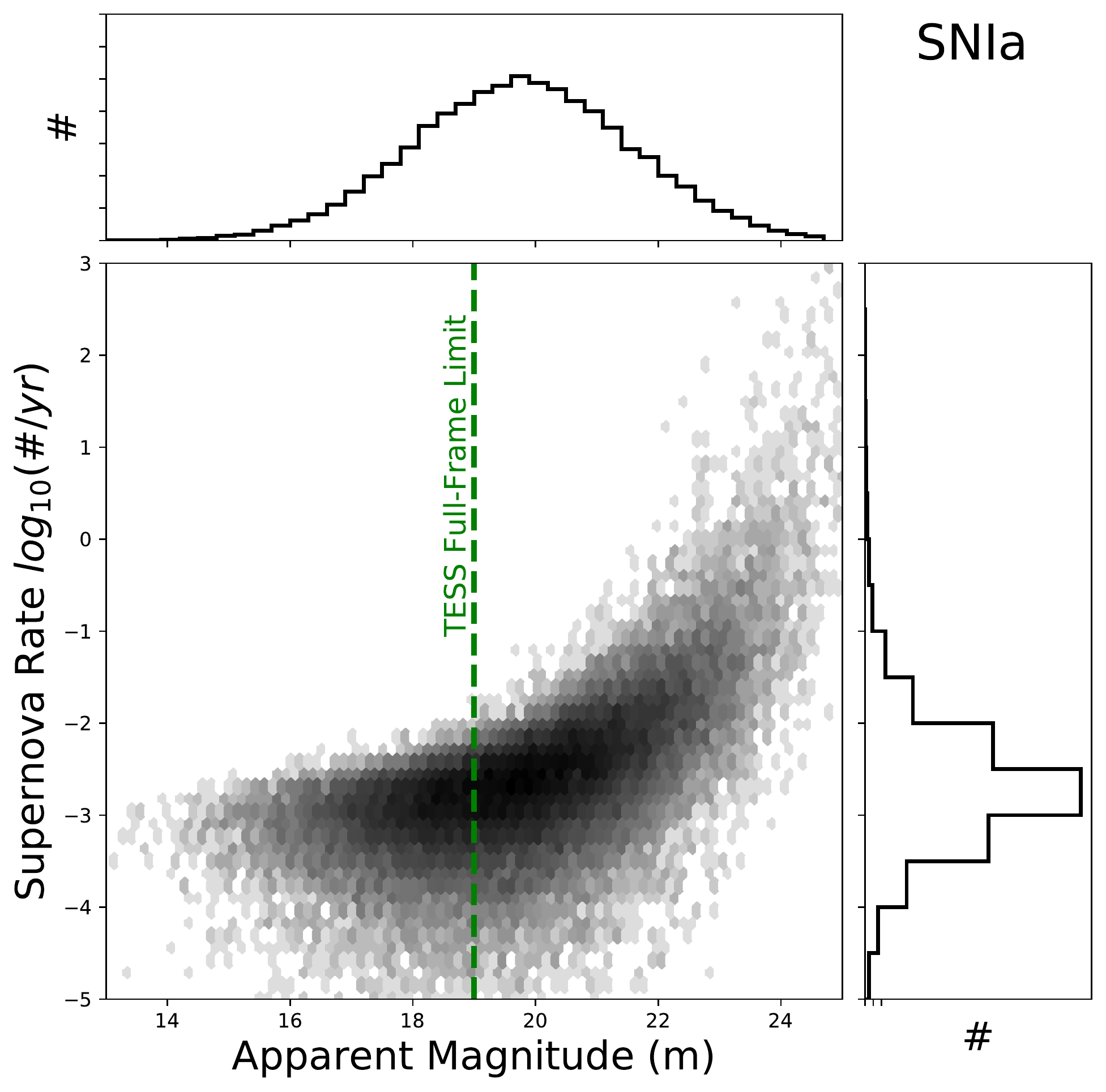}
  	\includegraphics[width=0.49\textwidth]{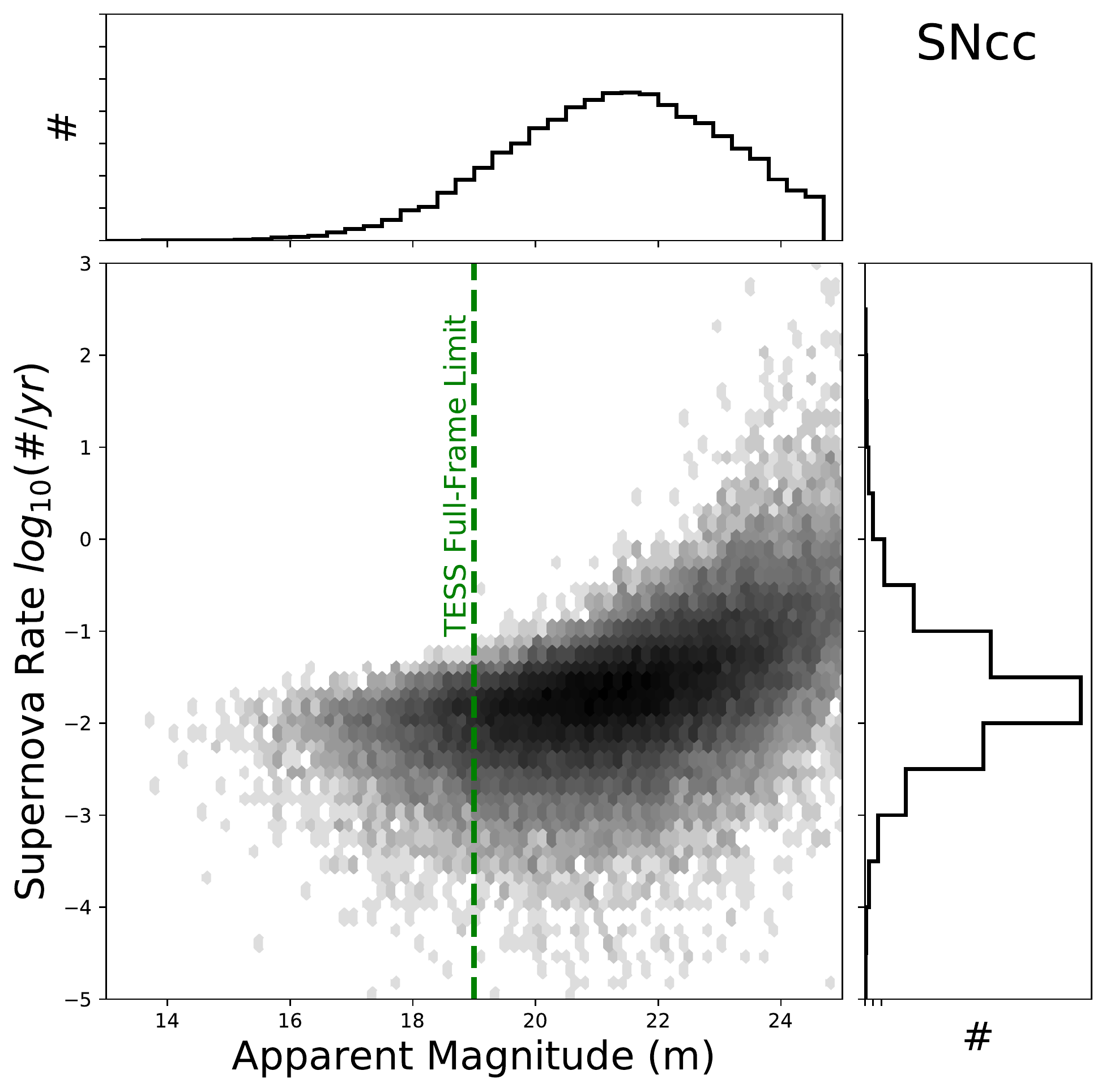}
	\caption{\label{f:3plot} The distribution of apparent magnitudes and supernova rates for an all-sky distribution of strong lenses for both Type Ia and core-collapse supernovae. The parameter space covered is either highly magnified and rare, or common and low magnification. }
  \end{center} 
\end{figure*}

\begin{figure}
  \begin{center}
  	\includegraphics[width=0.5\textwidth]{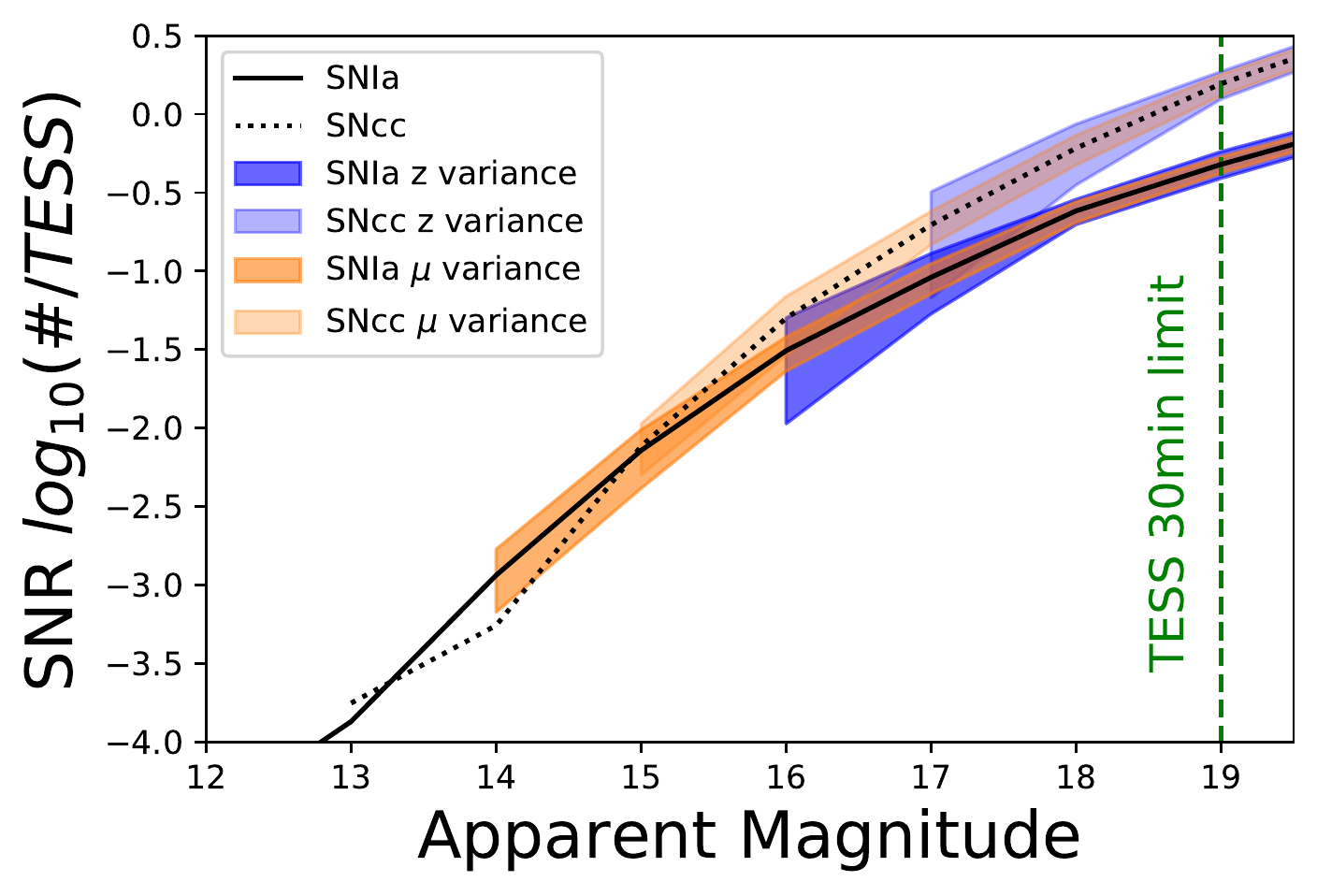}
	\caption{\label{f:m_snr}The summed all-sky supernova rate in gravitationally lensed host galaxies as a function of apparent magnitude in bins of 0.5mag. The figure shows the variance in the redshift distribution of the magnification distribution by 0.1 in peak or tail of the lognornormal distribution.}
  \end{center} 
\end{figure}

Starting from the randomly drawn redshift, we populate the target population in both apparent magnitude and the supernova rate. This is visualized in Figure \ref{f:3plot} for both types of supernova under consideration. The rates of occurrence of SNIa that are being lensed by a massive galaxy along the line of sight is relatively low, but because these are brighter to start with (Figure \ref{f:apparent-magnitude}), a sizeable number make it across the fiducial TESS detection limit. 
The core-collapse supernovae start almost two magnitudes dimmer than the SNIa, but are much more common. Their higher occurrence rate pushes them into competitive numbers above the TESS detection limit.

To find the total number of lensed supernovae of the two types --SNIa and SNcc-- we sum the occurrence rates in Figure \ref{f:3plot} and show these in Figure \ref{f:m_snr}. Only a few supernovae of both kinds over the whole sky are close enough and magnified enough by the gravitational lenses to be detectable with TESS: 
 16.37 SNIa/year and 44.1 SNcc/year. However, that is a reasonable occurrence rate to start looking for a signal in the TESS data of the first two years. 

TESS does not monitor the whole sky continuously, but there is considerable overlap between its campaigns as it covers 1/28th of the sky at a time in 27-day periods, completing the full sky in two years. Two ``continuous viewing zones'' are on each Ecliptic pole, resulting in 357 day coverage of each. There is some overlap between other sectors as well. 
To first order, this means 1.3 lensed SNIa and 13.4 lensed SNcc are in the TESS primary mission data (first two years). This drops to 0.5 SNIa and 6 SNcc if we limit ourselves to $m<18$ to ensure a good light curve fit. Note that this all-sky estimate was arrived at assuming SLACS-like lenses with source redshifts (Figure \ref{f:z}) and magnifications (Figure \ref{f:magnification}) and higher redshift and magnification events are excluded, underestimating the rate. 

In the extended mission of TESS, one could consider optimizing the detection rate by TESS of these lensed supernovae: longer campaigns to ensure the peak of the light curve is observed, an alert system based on rising sources, and  full-frame detections of a supernova.

\section{Why are lensed SNIa so Important?}

The occurrence of a supernova in a lensed system is rare, but these are significant opportunities for an independent test of our current understanding of cosmology. Multiple images of the background source galaxy each carry a separate image of the supernova, with a time delay between our observation of each supernova image that is expected to be on the order of days. The exact timing of each image of the supernovae brightening tells us the difference in the length of the path the light took, an independent test of General Relativity, Dark Matter and Dark Energy. 
Two supernovae in the source of a strongly lensed galaxy have been reported \citep{Quimby14,Goobar17}. 
Only one supernova with multiple Hubble imaging has been observed \citep[SN Refsdahl][]{Kelly15a}, but this {\em one object}, well characterized, already constrained the universe's cosmology \citep{Vega-Ferrero18, Grillo18, Williams19,Pierel19}.

\begin{figure}
    \centering
    \includegraphics[width=0.49\textwidth]{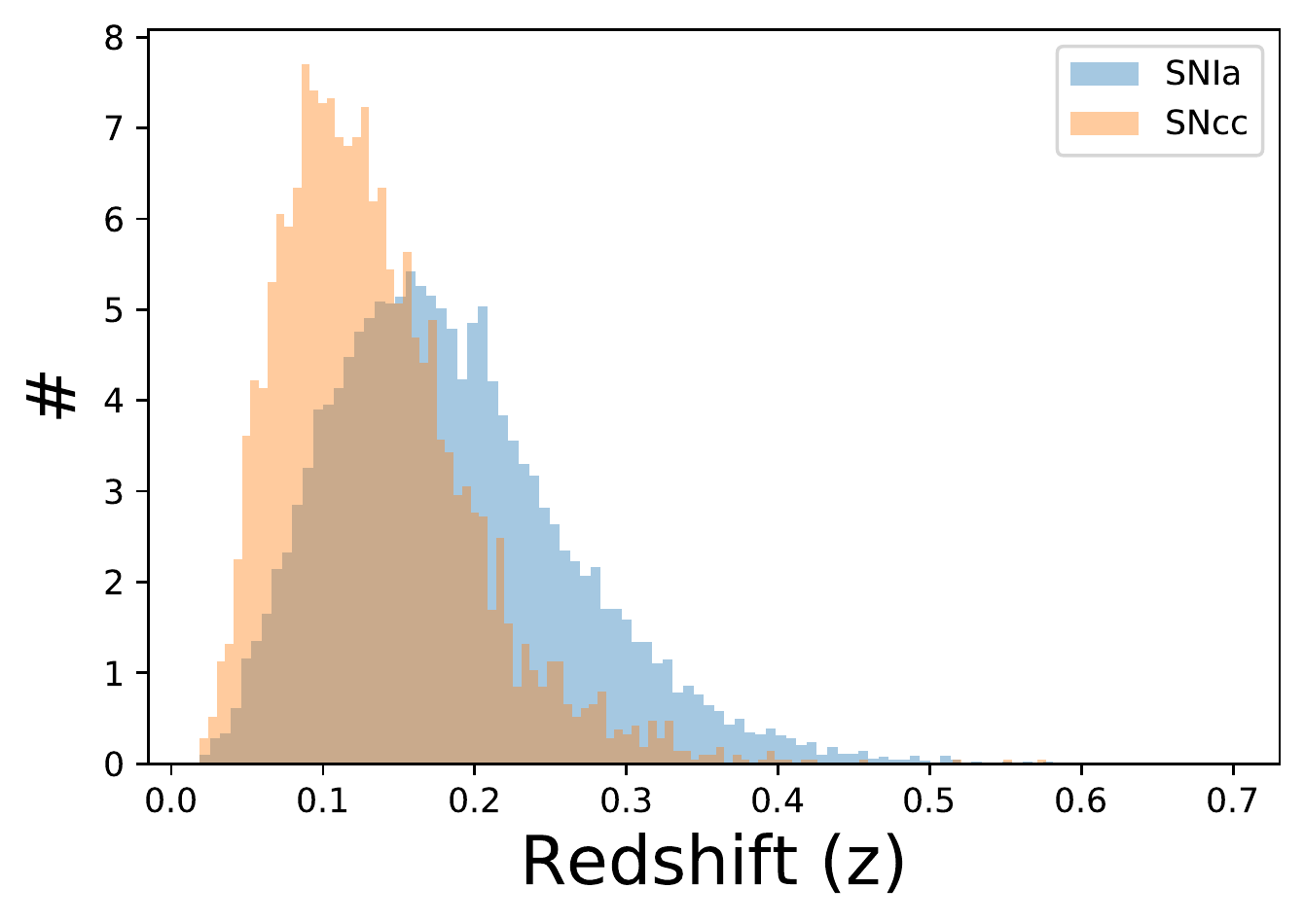}
    \caption{The distribution of SN, SNIa and SNcc that could be detected by TESS ($m<19$). Both peak before the assumed redshift distribution does (Figure \ref{f:z}). The source redshift distribution as found from the spectroscopically identified is a good proxy for the redshift distribution of sources whose supernova could be detected with TESS.  }
    \label{f:SN:selection:z}
\end{figure}
\section{Discussion}

We made several assumptions in the estimate of the observable supernovae in lensed systems with TESS: \begin{enumerate}
    \item The strong lenses identified in  \cite{Petrillo19} are representative and complete for all strong lenses in the sky. This is most likely a complete survey of the most massive and closest lens galaxies (and therefore with most clearly identifiable arcs) but unlikely to be a complete census as more blended systems are inherently missed by the machine learning algorithm. Their on-sky density of $\sim0.8/deg^2$ is therefore an {\em underestimate}. We adopted the more complete value of $1.27/deg^2$ from \cite{Knabel20} to estimate the observation rates. 
    \item The magnification distribution for the on-sky strong lenses is similar to the distribution found by \cite{Shu18} for the spectroscopically identified lenses, which are limited in diameter on the sky by the spectroscopic fiber of the survey and therefore favor slightly more distant or lower-mass galaxy lenses. \cite{Petrillo19} show their sample overlaps well with SLACS in mass-redshift space and is therefore likely representative. Thus, the magnifications assumed are likely an {\em underestimate} of the distribution of all on-sky lenses. 
    \item The source galaxy redshift distribution from \cite{Holwerda15} and \cite{Shu18} combined is representative of the ones observed all over the sky. These spectroscopically identified lenses are biased towards more distant lenses and hence sources. The redshift distribution may therefore be a slight {\em overestimate} of the actual on-sky redshift values (cf Figure \ref{f:SN:selection:z}). However, higher redshift but strongly magnified events are not considered and this results in a net underestimate of the predicted rate. 
    \item The supernova rates were consequently as high as those in \cite{Shu18} for star-forming galaxies at higher redshifts. The lensed arcs in \cite{Petrillo19} are selected by their blue colour and are therefore likely also star-forming galaxies, but --as pointed out above-- at lower redshifts. The SNR may therefore be an {\em overestimate} thanks to the possible overestimate of the redshift distribution.
   \item The host galaxy attenuation curve is for a smaller star-forming galaxy \citep[Figure \ref{f:host-attenuation},][]{Holwerda15c} but supernovae occur in all kinds of galaxies. We assume this is a good approximation of the source galaxies but it may constitute an underestimate of the SNae's attenuation from the host galaxy.
   \item We have ignored overlap in TESS field coverage (e.g. ecliptic poles) for the estimated all-sky rates. In effect this will improve TESS's odds since there is substantial field overlap (especially in the Northern campaign, see section \ref{s:polemonitoring}).
\end{enumerate}

Three of our six assumptions cause us to {\em underestimate} the number of observable supernovae in the TESS survey. The overestimate of the supernova rate in the source galaxies and smaller volume may lower the number. Hence, we treat our (approximate) estimate of the lensed supernovae rate as an underestimate for the TESS observations.

Figure \ref{f:m_snr} shows the variance in the distribution of SNR and apparent magnitude if we vary the redshift distribution in both center and width (Figure \ref{f:z}). Figure \ref{f:m_snr} shows the same variance for changing the distribution of magnifications (Figure \ref{f:magnification}). Variance in the redshift or magnification distribution are not substantially different from Figure \ref{f:m_snr} and only show stochasticity at the brightest apparent magnitudes. The greatest variance in the distribution occurs with the change in redshift distribution of the source galaxies. This distribution is difficult to predict, which is why we opted for a simple lognormal based on existing data to approximate the distribution observed in \cite{Holwerda15} and \cite{Shu18}. 

The innovation in strong lensing statistics is happening thanks to machine learning identification of strong lenses in imaging surveys. Recent efforts on such as the KiDS \citep{Petrillo17,Petrillo18,Petrillo19} and DES  \citep{Jacobs19,Huang20a,Huang20b} surveys have increases the numbers of known lenses from couple hundred to several thousand. These machine learning identifications prefer high-mass and closer lens galaxies, making the source arcs more easily identifiable in ground-based images. Knabel et al. \textit{in prep.} confirm several with existing GAMA spectroscopy showing weak second redshift signal because the arcs fall mostly outside the spectroscopic fiber aperture. These are ideal systems for TESS to monitor as their magnification is higher and sources are well separated from the lensing galaxy.
The on-sky density of these strong galaxy-galaxy lenses go a long way in explaining why a relatively low-redshift supernovae such as iPTF16geu \citep{Goobar17} can be found with high magnification (z=0.2 lens). \\
 
\begin{figure*}
  \begin{center}
  	\includegraphics[width=\textwidth]{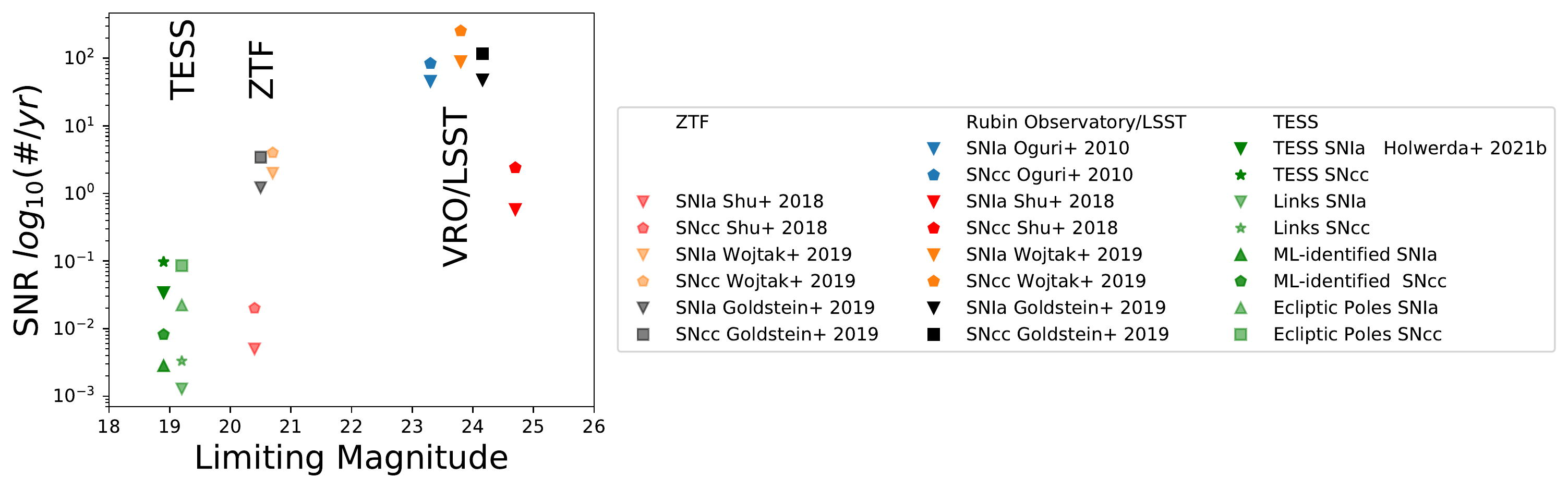}
	\caption{\label{f:snr:surveys}The detection rates for different lensed supernovae types for the ZTF and Rubin Observatory/LSST surveys from \protect\cite{Oguri10a}, \protect\cite{Wojtak19}, \protect\cite{Goldstein19} and \protect\cite{Shu18}. 
	The green points are different estimates using TESS as the survey instrument: all-sky TESS detections of lensed supernovae, monitoring the LiNKS Machine-learning identified strong lensing systems, monitoring all machine learning identified strong lensing systems in DES/KiDS/SLACS etc, and the continuous monitoring of the Ecliptic poles. 
	The TESS all-sky and the Ecliptic Poles 
	blind searches with TESS may be competitive in the near-term for all-sky searches for lensed supernovae and will remain competitive in the ecliptic poles thanks to continuous coverage. }
  \end{center} 
\end{figure*}

Supernovae rates in gravitationally magnified systems are presented in \cite{Oguri10a}, \cite{Shu18}, \cite{Wojtak19}, and \cite{Goldstein19} with a variety of assumptions and for different time-domain surveys.
These estimates for ZTF and Rubin Observatory/LSST are shown in Figure \ref{f:snr:surveys} for reference. The majority are for blind searches and a few for monitoring well-known strong lensing samples. TESS offers a different approach, monitoring large samples of candidate lenses as well as all-sky or polar blind searches.


Any TESS detection in a lensed system offers the possibility to repeat the supernova timing experiment conducted by \citep{Kelly15a} in a single lensing galaxy instead of a lensing galaxy cluster. Current estimates are for the near-future Vera Rubin Observatory, but even for that powerful transient observatory, supernovae in lensed systems observation rates are of order unity \citep{Goldstein19}. That makes the TESS observations potentially competitive. 


\section{Conclusion}

TESS is promising to be an amazing tool for a wide variety of astronomical topics, ranging from the exoplanets it was meant to find to stellar seismology and other transient phenomena. We present here the odds of not only finding a Supernova --several have already been discovered by TESS in combination with the All-Sky Automated Survey for Supernovae (ASAS-SN) see e.g. \cite{Vallely19}-- but an estimate of how many of these have been significantly lensed by a strong gravitational lens of a foreground massive galaxy.

The total number of lensed SNIa and SNcc per year in the TESS visibility envelope is proportionally lower, resulting in about 0.5 or 1.3 SNIa and 6 or 13.4 SNIcc potentially identifiable (assuming a $m<18$ or $m<19$ limiting magnitude) in the TESS primary 2-year mission. 
Alternatively, one could monitor the KiDS-identified strong lensing systems alone \citep[LiNKS][]{Petrillo19}. This would lower the expected rates by another order of magnitude in exchange for the certainty that these are lensed supernovae.
The odds of finding one each year are approximately 18\% and 43\% for SNIa and SNcc respectively and 0.6\% and 1.3\% SNIa and SNcc per year monitoring known lenses. 
With the results from the DES search for strong lenses \citep{Jacobs19,Huang20a,Huang20b}, the total number of known strong lensing galaxies, mostly found through machine learning and worth monitoring is close to 3000, doubling the numbers for just LiNKS (see Figure \ref{f:snr:surveys}).

Figure \ref{f:snr:surveys} shows the TESS and KiDS-monitored supernova rates in comparison to the rates predicted for the Zwicky Transient Facility and the Vera C. Rubin Observatory by \cite{Goldstein19} and \cite{Shu18}. These are for the full survey \citep{Goldstein19} or monitoring known strong lenses \citep{Shu18}, similar to our proposed TESS and KiDS-identified \citep[by][]{Petrillo19} strong lenses. 
The TESS numbers are an order of magnitude below the other surveys' expected supernova yields. We note the all-sky TESS yield is comparable with the ZTF dedicated lens monitoring (Figure \ref{f:snr:surveys}).
Monitoring known lenses such as those in the KiDS survey \citep[LinKS, lenses in a square degree][]{Petrillo19}, similarly yields an order of magnitude fewer supernovae for TESS. One viable way to improve TESS (and other transient observatories) performance is to increase the number of known lens systems to monitor. Given that strong lens selection thusfar has used aperture-limited spectra with a blended signal from both galaxies, there is a substantial yield of lower-redshift, strongly lensing galaxies left to find \citep[see][for a discussion on different lensing detection methods]{Knabel20} and a strong motivation for machine learning identification applied on all-sky surveys. Monitoring machine learning identified lensing systems (some 3000 now in total from KiDS and DES) is as effective as the entire all-sky campaign of TESS (Figure \ref{f:snr:surveys}). 

The odds of finding a supernova with TESS are low for its primary all-sky mission, but the high cadence (30 min) and long campaign of TESS (27 days) would result in accurate light curves, even if these were a mix of two light curves observed in different lensed images of the source galaxy. The possibility of observing such events and the potential pay-offs of an independent Hubble constant measurement could make this a worthy additional science to be conducted with the TESS telescope during its extended mission. Considering the science potential for an extended mission, the potential number of supernovae to be discovered goes up commensurately, especially when a modified observing strategy is followed with longer campaigns on each sector and faster processing of each sector allows rapid successful spectroscopic and high-resolution imaging follow-up of potential supernovae. 

As a first step, the existing TESS archive can be scoured for the signal of a multiple-lensed supernova with confirmation using multicolour ground-based surveys (e.g. SDSS or DES). If this proof of concept works, a rapid pipeline for detection of such rare events should be a priority for timely follow-up; high resolution imaging to discriminate each SN image and spectroscopy to confirm supernova type. 

Alternatively, the number of strong lensing galaxies which are candidates for TESS monitoring will be continuously increased. This can be additional motivation for Machine Learning efforts to find strong lensing galaxies at low redshift in all-sky imaging such as the DES search \citep{Jacobs19,Huang20a,Huang20b}. 

\begin{figure}
    \centering
    \includegraphics[width=0.5\textwidth]{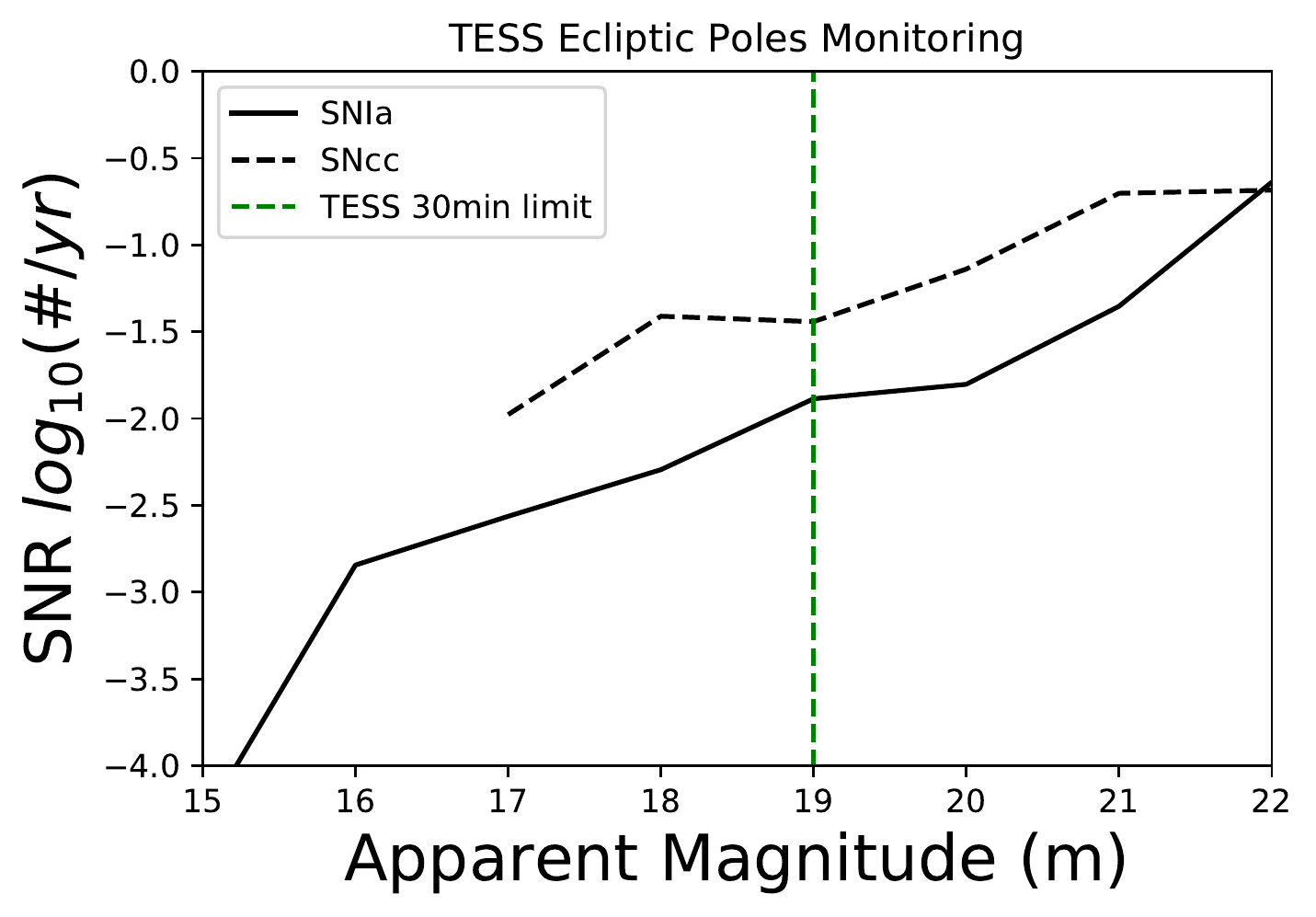}
    \caption{The estimate of number of lensed supernovae in the ecliptic poles, one of which is monitored for 357 days out of the year with TESS. }
    \label{f:poles}
\end{figure}

\subsection{Ecliptic Pole Monitoring}
\label{s:polemonitoring}
We argue that special attention should be given to the Ecliptic Poles where TESS has continuous coverage and JWST continuous viewing zone lies (always available for rapid follow-up). These regions are already of intense interest for transient monitoring, especially the Northern Ecliptic pole is promising for extra-galactic work \citep[see][]{Jansen18}. 

TESS has an undeniable advantage over ground-based surveys in these regions thanks to the near year-round monitoring of one of these poles. Figure \ref{f:poles} shows the number of lensed supernovae one can expect in these 60 deg$^2$. This estimate is much simpler as TESS observed any particular pole 50\% of the time, improving the odds of observing one: a 2\% chance of supernova type Ia and 4\% chance of a core-collapse supernova each year, both in a lensed galaxy and observable with TESS ($m<19$), assuming 1.2 lensing system per square degree (Figure \ref{f:snr:surveys}). An added benefit is that the complete lightcurve of the supernova is likely to be fully sampled by TESS alone.

In the South, much of the ecliptic pole is crowded by the Large Magellanic Cloud but the Northern Ecliptic pole offers both a reasonable chance of success and JWST continuous follow-up potential. With the Northern pole unavailable for the Rubin Observatory, this gives TESS an unique parameter space for a potential, near-future and high scientific return science target for its extended mission.

\section*{Data Availability}

The data underlying this article are available in the article and in its online supplementary material (jupyter notebook).

\section*{Acknowledgements}

The material is based upon work supported by NASA Kentucky under NASA award No: NNX15AR69H.
This research has made use of the NASA/IPAC Extragalactic Database (NED) which is operated by the Jet Propulsion Laboratory, California Institute of Technology, under contract with the National Aeronautics and Space Administration. 
%
%
This research has made use of NASA's Astrophysics Data System.
This research made use of Astropy, a community-developed core Python package for Astronomy \citep{Astropy-Collaboration13a} and matplotlib, a Python library for publication quality graphics \citep{Hunter07}. PyRAF is a product of the Space Telescope Science Institute, which is operated by AURA for NASA. This research made use of SciPy \citep{scipy}. 

%

\bsp	
\label{lastpage}
\end{document}